\documentclass[conference]{IEEEtran}
\IEEEoverridecommandlockouts

\usepackage{cite}

\ifCLASSINFOpdf
\usepackage[pdftex]{graphicx}
\else
\usepackage[dvips]{graphicx}
\fi
\usepackage{amsmath}
\usepackage{multirow}
\usepackage{color}
\usepackage{amsfonts}
\usepackage[normalem]{ulem}
\usepackage{pgfplots}
\pgfplotsset{compat=1.8}
\usepackage{filecontents} 

\begin{document}
%
\title{A Digital Neuromorphic Architecture Efficiently Facilitating Complex Synaptic Response Functions Applied to Liquid State Machines}



%
\author{\IEEEauthorblockN{Michael R. Smith\IEEEauthorrefmark{1},
Aaron J. Hill\IEEEauthorrefmark{1},
Kristofor D. Carlson\IEEEauthorrefmark{1},
Craig M. Vineyard\IEEEauthorrefmark{1},
Jonathon Donaldson\IEEEauthorrefmark{1},\\
David R. Follett\IEEEauthorrefmark{2},
Pamela L. Follett\IEEEauthorrefmark{2}\IEEEauthorrefmark{3},
John H. Naegle\IEEEauthorrefmark{1},
Conrad D. James\IEEEauthorrefmark{1} and
James B. Aimone\IEEEauthorrefmark{1}}
\IEEEauthorblockA{\IEEEauthorrefmark{1}Sandia National Laboratories,
Albuquerque, NM 87185 USA\\
Email:\{msmith4, ajhill, kdcarls, cmviney, jwdonal, jhnaegl, cdjame, jbaimon\}@sandia.gov}
\IEEEauthorblockA{\IEEEauthorrefmark{2}Lewis Rhodes Labs,
Concord, MA 01742 USA\\
Email:\{drfollett, plfollett\}@earthlink.net}
\IEEEauthorblockA{\IEEEauthorrefmark{3}Tufts University, Medford, MA 02155 USA}
\thanks{This work was supported by Sandia National Laboratories’ Laboratory
Directed Research and Development (LDRD) Program under the Hardware
Acceleration of Adaptive Neural Algorithms (HAANA) Grand Challenge project.
Sandia National Laboratories is a multi-mission laboratory managed and
operated by Sandia Corporation, a wholly owned subsidiary of Lockheed
Martin Corporation, for the U. S. Department of Energy’s National Nuclear
Security Administration under Contract DE-AC04-94AL85000.}}


\maketitle

\begin{abstract}
	Information in neural networks is represented as weighted connections, or synapses, between neurons.
	This poses a problem as the primary computational bottleneck for neural networks is the vector-matrix multiply when inputs are multiplied by the neural network weights.
	Conventional processing architectures are not well suited for simulating neural networks, often requiring large amounts of energy and time.
	Additionally, synapses in biological neural networks are not binary connections, but exhibit a nonlinear response function as neurotransmitters are emitted and diffuse between neurons.
	Inspired by neuroscience principles, we present a digital neuromorphic architecture, the {\it Spiking Temporal Processing Unit} (STPU), capable of modeling arbitrary complex synaptic response functions without requiring additional hardware components.
	We consider the paradigm of spiking neurons with temporally coded information as opposed to non-spiking rate coded neurons used in most neural networks.
	In this paradigm we examine liquid state machines applied to speech recognition and show how a liquid state machine with temporal dynamics maps onto the STPU---demonstrating the flexibility and efficiency of the STPU for instantiating neural algorithms.
\end{abstract}


%
\IEEEpeerreviewmaketitle

\section{Introduction}
\label{section:intro}
Neural-inspired learning algorithms are achieving state of the art performance in many application areas such as speech recognition \cite{Deng2013_ICASSP}, image recognition \cite{Ciresan2012_CVPR}, and natural language processing \cite{Socher2013_EMNLP}.
Information and concepts, such as a dog or a person in an image, are represented in the synapses, or weighted connections, between the neurons.
The success of a neural network is dependent on training the weights between the neurons in the network.
However, training the weights in a neural network is non-trivial and often has high computational complexity with large data sets requiring long training times.

One of the contributing factors to the computational complexity of neural networks is the vector-matrix multiplications (the input vector multiplied by the synapse or weight matrix).
Conventional computer processors are not designed to process information in the manner that a neural algorithm requires (such as the vector-matrix multiply).
Recently, major advances in neural networks and deep learning have coincided with advances in processing power and data access.
However, we are reaching the limits of Moore's law in terms of how much more efficiency can be gained from conventional processing architectures.
In addition to reaching the limits of Moore's law, conventional processing architectures also incur the von Neumann bottleneck \cite{Backus1978_CommACM} where the processing unit's program and data memory exist in a single memory with only one shared data bus between them.

In contrast to conventional processing architectures which consist of a powerful centralized processing unit(s) that operate(s) in a mostly serialized manner, the brain is composed of many simple distributed processing units (neurons) that are sparsely connected and operate in parallel.
Communication between neurons occurs at the synaptic connection which operate independently of the other neurons that are not involved in the connection. 
Thus, vector-matrix multiplications are implemented more efficiently facilitated by parallel operations.
Additionally, the synaptic connections in the brain are generally sparse and information is encoded in a combination of the synaptic weights and the temporal latencies of a spike on the synapse \cite{Follett2009_JofChildNeur}. 
Biological synapses are not simply a weighted binary connection but rather exhibit a non-linear synaptic response function due to the release and dispersion of neurotransmitters in the space between neurons.

\begin{figure*}[!t]
\centering
\includegraphics[width=6in]{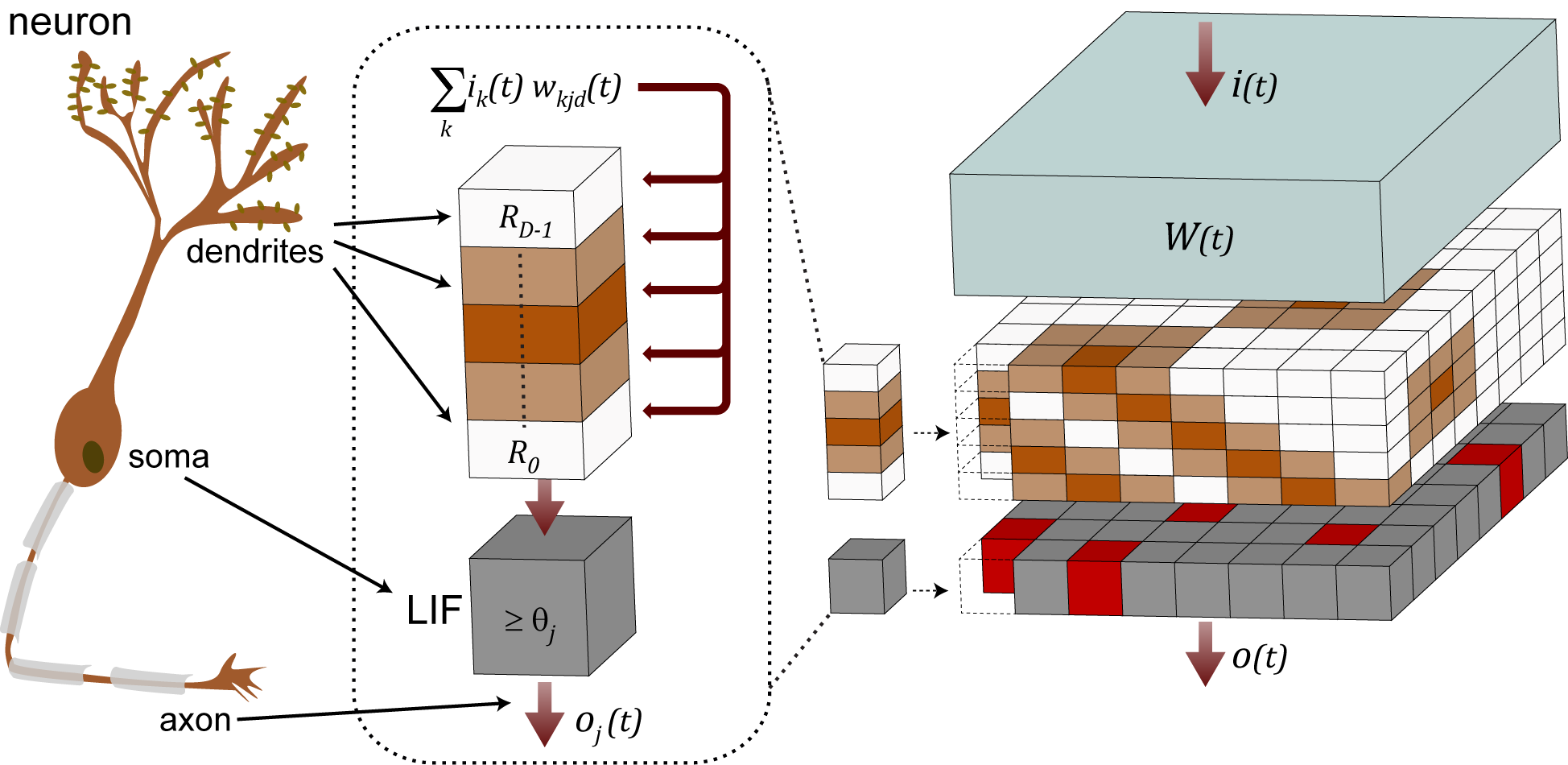}
\caption{High level overview of the STPU. The STPU is composed of a set of leaky integrate and fire neurons. Each neuron has an associated temporal buffer such that inputs can be mapped to a neuron with a time delay. $\mathbf{W}(t)$ is the neuronal encoding transformation which addresses connectivity, efficacy and temporal shift. The functionality of the STPU mimics the of functionality of biological neurons.}
\label{fig:STPU}
\end{figure*}

Biological neurons communicate using simple ``data packets,'' that are generally accepted as binary {\it spikes}.
This is in contrast to the neuron models used in traditional artificial neural networks (ANN) which are commonly rate coded neurons.
Rate coded neurons encode information between neurons as a real-valued magnitude of the output of a neuron---a larger output represents a higher firing rate.
The use of rate coded neurons stems from the assumption that the firing rate of a neuron is the most important piece of information, whereas temporally coded neurons encode information based on when a spike from one neuron arrives at another neuron.
Temporally coded information has been shown to be more powerful than rate coded information and more biologically accurate \cite{Sejnowski1995_Nature} . 


Based on these neuroscience principles, we present the {\it Spiking Temporal Processing Unit} (STPU), a novel neuromorphic hardware architecture designed to mimic neuronal functionality and alleviate the computational restraints inherent in conventional processors.
Other neuromorphic architectures have shown very strong energy efficiency \cite{Merolla2014_Science}, powerful scalability \cite{Furber2013_TransOnComp}, and aggressive speed-up \cite{Schemmel2008_IJCNN} by utilizing the principles observed in the brain.
We build upon these efforts leveraging the benefits of low energy consumption, scalability, and run time speed ups and include an efficient implementation of arbitrarily complex synaptic response functions in a digital architecture.
This is important as the synaptic response function has strong implications in spiking recurrent neural networks \cite{Zhang2015_TNNLS}.

We also examine liquid state machines (LSMs) \cite{Maass2002_NeuralComput} to show how the constructs available in the STPU facilitate complex dynamical neuronal systems.
While we examine the STPU in the context of LSMs, the STPU is a general neuromorphic architecture.
Other spiked-based algorithms have been implemented on the STPU \cite{Verzi2017_IJCNN, Verzi2016_NN}.

In Section \ref{section:STPU}, we present the STPU.
A high level comparison with other neuromorphic architectures is presented in Section \ref{section:comp}.
We present LSMs in Section \ref{section:LSM}.
In Section \ref{section:mapping}, we examine how LSMs map onto the STPU and show results from running the LSM on the STPU.
We conclude in Section \ref{section:conc}.

\section{The Spiking Temporal Processing Unit} 
\label{section:STPU}

In this section, we describe the {\it Spiking Temporal Processing Unit} (STPU) and how the components in the STPU map to functionality in biological neurons.
The design of the STPU is based on the following three neuroscience principles observed in the brain:
1) the brain is composed of simple processing units (neurons) that operate in parallel and are sparsely connected,
2) each neuron has its own local memory for maintaining temporal state, and
3) information is encoded in the connectivity, efficacy, and signal propagation characteristics between neurons.
A high-level overview of a biological neuron and how its components map onto the STPU are shown in Figure \ref{fig:STPU}.
The STPU derives its dynamics from the leaky integrate and fire (LIF) neuron model \cite{Dayan2001_TheoreticalNeuroSci}.
Each LIF neuron $j$ maintains a membrane potential state variable, $v_j$, that tracks its stimulation at each time step based on the following differential equation \cite{Zhang2015_TNNLS}:
\begin{equation}
\frac{dv_j}{dt} = - \frac{v_j}{\tau_j} + \sum_k \sum_l w_{kj} \cdot s(t-t_{kl}-\Delta_{kl}).
\label{eq:LIF}
\end{equation}
The variable $\tau_j$ is the time constant of the first-order dynamics, $k$ is the index of the presynaptic neuron, $w_{kj}$ is the weight connecting neuron $j$ to neuron $k$, $t_{kl}$ is the time of the $l^{\text{th}}$ spike from neuron $k$, $\Delta_{kl}$ is the synaptic delay from neuron $k$ on the $l^{\text{th}}$ spike, and $s(\cdot)$ is the dynamic synaptic response function to an input spike.
In the LIF model, neuron $j$ will fire if $v_j$ exceeds a threshold $\theta_j$.
The synapses between input neurons to destination neurons are defined in the weight matrix $\mathbf{W}(t)$ for a given time $t$ as the weights between inputs and neurons can change over time.

Unique to the STPU, each LIF neuron has a local temporal memory buffer $R$ composed of $D$ memory cells to model synaptic delays.
When a biological neuron fires, there is a latency associated with the arrival of the spike at the soma of the postsynaptic neuron due to the time required to propagate down the axon of the presynaptic neuron and the time to propagate from the dendrite to the soma of the postsynaptic neuron ($\Delta_{kl}$).
The temporal buffer represents different synaptic junctions in the dendrites where a lower index value in the temporal buffer constitutes a dendritic connection closer to the soma and/or a shorter axon length than one with a larger index value.
Thus, synapses in the STPU are specified as a weight $w_{kjd}$ from a source input neuron $k$, to a destination neuron $j$ in the $d^{\text{th}}$ cell of the temporal buffer, $d\in{\{0, 1, 2,\dots, D-1\}}$. 
This allows multiple connections between neurons with different synaptic delays.
At each time step a summation of the product of the inputs $\mathbf{i}(t)$ and synaptic weights $\mathbf{W}(t)$ occurs and is added to the current value in that position of the temporal buffer $\hat{R}_d(t) = R_d(t) + \sum_k i_k(t) w_{kjd}(t)$ where $\hat{R}(t)$ is a temporary state of the temporal buffer.
The value in each cell of the temporal buffer is then shifted down one position, that is $R_d(t+1) = \hat{R}_{d-1}(t)$.
The values at the bottom of the buffer are fed into the LIF neuron.

\begin{figure}
	\centering
	\includegraphics[width=3in]{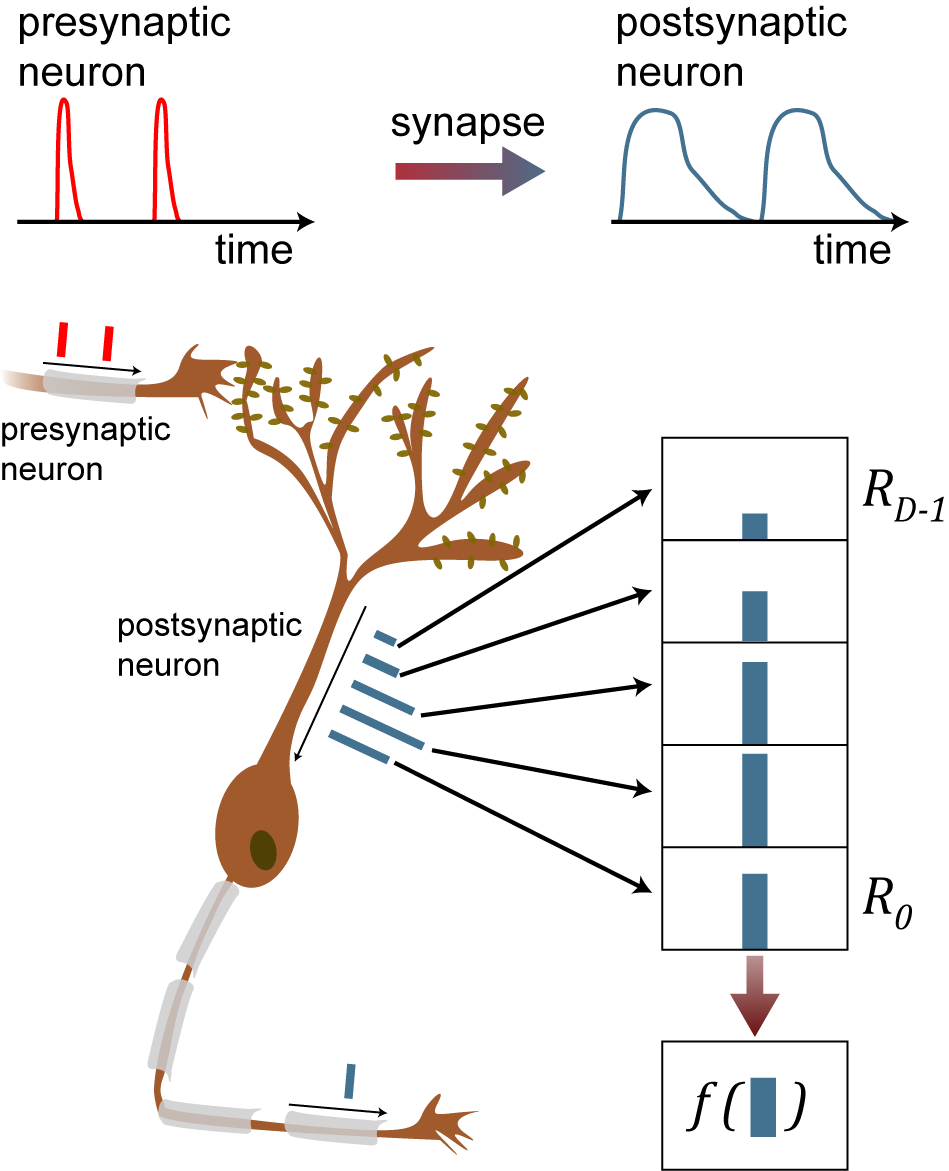}
	\caption{Spike propagation along the axon and across the synapse.
		The spike propagated on the axon is generally accepted as a binary spike. 
		Upon arrival at the synapse, the spike initiates a chemical reaction in the synaptic cleft which stimulates the postsynaptic neuron.
		This chemical reaction produces an analog response that is fed into the soma of the postsynaptic neuron.
		In the STPU, arbitrary synaptic response functions are modeled efficiently using the temporal buffer.
		The synaptic response function is discretely sampled and encoded into the weights connecting one neuron to another and mapped to the corresponding cells in the temporal buffer.}
	\label{fig:synapticResponse}
\end{figure}

In biological neurons, when a neuron fires a (near) binary spike is propagated down the axon to the synapse, which defines a connection between neurons.
The purpose of the synapse is to transfer the electric activity or information from one neuron to another neuron.
Direct electrical communication does not take place, rather a chemical mediator is used.
In the presynaptic terminal, an action potential from the emitted spike causes the release of neurotransmitters into the synaptic cleft (space between the pre and postsynaptic neurons) from the synaptic vescles. 
The neurotransmitters cross the synaptic cleft and attach to receptors on the postsynaptic neuron injecting a positive or negative current into the postsynaptic neuron.
Through a chemical reaction, the neurotransmitters are broken down in receptors on the postsynaptic neuron and are released back into the synaptic cleft where the presynaptic neuron reabsorbs the broken down molecules to synthesize new neurotransmitters.
In terms of electrical signals, the propagation of activation potentials on the axon is a digital signal as shown in Figure \ref{fig:synapticResponse}.
However, the chemical reactions that occur at the synapse to release and reabsorb neurotransmitters are modeled as an analog signal.

\begin{table*}[!t]
	\renewcommand{\arraystretch}{1.3}
	\caption{High-level Comparison of the STPU with True North and SpiNNaker.}
	\label{table:comp}
	\centering
	\begin{tabular}{|l||lll|}
		\hline
		Platform: & STPU & TrueNorth & SpiNNaker\\
		\hline
		Interconnect: & 3D mesh multicast\footnotemark & 2D mesh unicast & 2D mesh multicast \\
		Neuron Model: & LIF & LIF\footnotemark & Programmable\footnotemark\\
		Synapse Model: & Programmable\footnotemark & Binary & Programmable\footnotemark \\
		\hline
	\end{tabular}
	\break
	\vskip 0.01 cm
	\addtocounter{footnote}{-5}
	\footnotemark\footnotesize{The 3D mesh is enabled due to the temporal buffer available for each neuron in STPU.\hfill}
	\break
	\footnotemark \footnotesize{TrueNorth provides a highly programmable LIF to facilitate additional neural dynamics.\hfill}
	\break
	\footnotemark \footnotesize{SpiNNaker provides flexibility for the neuron model, however more complex biological models are more computationally expensive.\hfill}
	\break
	\footnotemark\footnotesize{The synapse model is programmable in the STPU via the temporal buffer by discretely sampling an arbitrary synapse model.\hfill}
	\break
	\footnotemark\footnotesize{As with the neuron model, SpiNNaker is optimized for simpler synaptic models. More complex synaptic models incur a cost in computational complexity.\hfill}
	\vskip -1mm
\end{table*}

The behavior of the synapse propagating spikes between neurons has important ramifications on the dynamics of the liquid. 
In Equation \ref{eq:LIF}, the synaptic response function is represented by $s(\cdot)$. 
Following Zhang et al. \cite{Zhang2015_TNNLS}, the Dirac delta function $\delta(\cdot)$ can be used as the synaptic response function and is convenient for implementation on digital hardware.
However, the Dirac delta function exhibits static behavior.
Zhang et al. show that dynamical behavior can be modeled in the synapse by using the first-order response to a presynaptic spike:
\begin{equation}
\frac{1}{\tau^s}e^{-\frac{t-t_{kl}-\Delta{kl}}{\tau^s}}\cdot H(t-t_{kl}-\Delta_{kl})
\label{eq:firstOrder}
\end{equation}
where $\tau^s$ is the time constant of the first-order response, $H(\cdot)$ is the Heaviside step function, and $1/\tau^s$ normalizes the first-order response function. 
The dynamical behavior can also be implemented using a second-order dynamic model for $s(\cdot)$:
\begin{equation}
\frac{1}{\tau^s_1 - \tau^s_2}(e^{-\frac{t-t_{kl}-\Delta{kl}}{\tau^s_1}}-e^{-\frac{t-t_{kl}-\Delta{kl}}{\tau^s_2}})\cdot H(t-t_{kl}-\Delta_{kl})
\label{eq:secondOrder}
\end{equation}
where $\tau^s_1$ and $\tau^s_2$ are the time constants for the second order response and $1/(\tau^s_1-\tau^s_2)$ normalizes the second-order dynamical response function.
Zhang et al. showed significant improvements in accuracy and the dynamics of the liquid when using these dynamical response functions.

Implementing exponential functions in hardware is expensive in terms of the resources needed to implement exponentiation.
Considering that the STPU is composed of individual parallel neuronal processing units, each neuron would need its own exponentiation functionality.
Including the hardware mechanisms for each neuron to do exponentiation would significantly reduce the number of neurons by orders of magnitude as there are limited resources on an FPGA.
Rather than explicitly implement the exponential functions in hardware, we use the temporal buffer associated with each neuron.
The exponential function is discretely sampled and the value at each sample is assigned a connection weight $w_{kjd}$ from the presynaptic neuron $k$ to the corresponding cell $d$ in the temporal buffer of the postsynaptic neuron $j$.
Thus, a single weighted connection between two neurons is expanded to multiple weighted connections between the same two neurons.
This is shown graphically in Figure \ref{fig:synapticResponse}.
The use of the temporal buffer allows for an efficient implementation of the digital signal propagation down the axon of a neuron or the analog signal propagation between neurons at the synapse.

\section{Comparison with Other Neuromorphic Architectures}
\label{section:comp}

The STPU is not the first neuromorphic architecture.
Four prominent neuromorphic architectures are IBM's TrueNorth chip \cite{Merolla2014_Science}, the Stanford Neurogrid \cite{Benjamin2014_IEEE}, the Heidelberg BrainScaleS machine \cite{Schemmel2010_SymposiumCircuitSystems} and the Manchester Spiking Neural Network Architecture (SpiNNaker) \cite{Furber2014_IEEE}.
The Stanford Neurogrid and the Heidelberg BrainScaleS are analog circuits while TrueNorth and SpiNNaker are digital circuits.
As the STPU is also a digital system, we will focus on a comparison with TrueNorth and SpiNNaker.

The TrueNorth chip leverages a highly distributed crossbar based architecture designed for high energy-efficiency composed of 4096 cores. 
The base-level neuron is a highly parametrized LIF neuron.
A TrueNorth core is a $256 \times 256$ binary crossbar where the existence of the synapse is encoded at each junction, and individual neurons assign weights to particular sets of input axons.
The crossbar architecture allows for efficient vector-matrix multiplication.
TrueNorth only allows for point-to-point routing.
Each of the 256 neurons on a core is programmed with a spike destination addressed to a single row on a particular core which could be the same core, enabling recurrence, or a different core.
The crossbar inputs are coupled via delay buffers to insert axonal delays.
A neuron is not natively able to connect to multiple cores or to connect to a single neuron with different temporal delays.
As a work around, a neuron is to be replicated within the same core and mapped to the different cores.
For multiple temporal delays between two neurons (such as those in the STPU), there is no obvious mechanism for an implementation \cite{Severa2016_NIPS_WS}.

SpiNNaker is a massively parallel digital computer composed of simple ARM cores with an emphasis on flexibility. 
Unlike the STPU and TrueNorth, SpiNNaker is able to model arbitrary neuron models via an instruction set that is provided to the ARM core. 
SpiNNAker is designed for sending large numbers of small data packages to many destination neurons. 
While SpiNNaker was designed for modeling neural networks, it could potentially be used more generally due to its flexibility. 

The STPU architecture falls in between the TrueNorth and SpiNNaker architectures.
The STPU implements a less parameterized LIF neuron than TrueNorth, however, its routing of neural spikes is more flexible and allows a multicast similar to SpiNNaker rather than the unicast used in TrueNorth.
A key distinguishing feature of the STPU is the temporal buffer associated with each neuron, giving the STPU 3-dimensional routing.
A high-level summary of the comparison of STPU with TrueNorth and SpiNNaker is shown in Table \ref{table:comp}.

\section{Liquid State Machines}
\label{section:LSM}

The liquid state machine (LSM) \cite{Maass2002_NeuralComput} is a neuro-inspired algorithm that mimics the cortical columns in the brain.
It is conjectured that the cortical microcircuits nonlinearly project input streams into a high-dimensional state space.
This high-dimension representation is then used as input to other areas in the brain where learning can be achieved.
The cortical microcircuits have a sparse representation and fading memory---the state of the microcircuit ``forgets'' over time.
While LSMs may be able to mimic certain functionality in the brain, it should be noted that LSMs do not try to explain how or why the brain operates as it does.

In machine learning, LSMs are a variation of recurrent neural networks that fall into the category of reservoir computing (RC) \cite{Lukosevicius2009_CSReview} along with echo state networks \cite{Jaeger2003_NIPS}.
LSMs differ from echo state machines in the type of neuron model used.
LSMs use spiking neurons while echo state machines use rate coded neurons with a non-linear transfer function.

\begin{figure}
	\centering
	\includegraphics[width=3in]{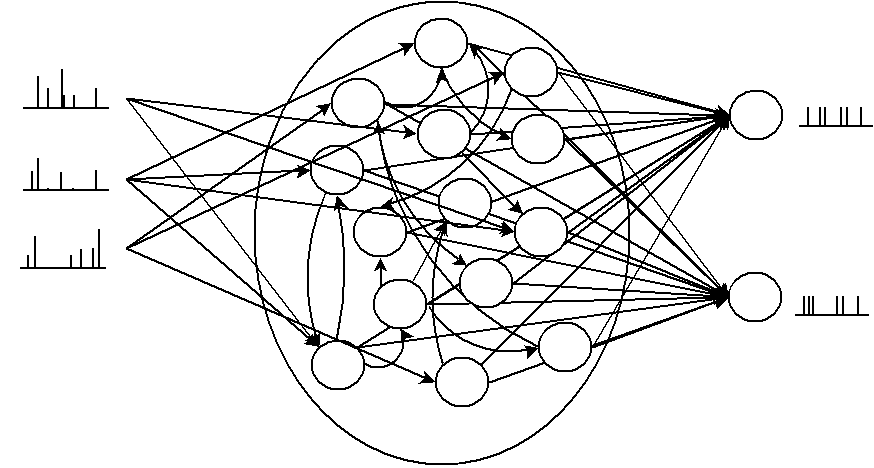}
	\caption{A liquid state machine, composed of three components: 1) a set of input neurons, 2) the liquid---a set of recurrent spiking neurons, and 3) a set of readout neurons with plastic synapses that can read the state of the neurons in the liquid.}
	\label{fig:LSM}
	\vskip -3mm
\end{figure}

LSMs operate on temporal data composed of multiple related time steps.
LSMs are composed of three general components:
1) input neurons,
2) randomly connected leaky-integrate and fire spiking neurons called the liquid, and
3) readout nodes that read the state of liquid.
A diagram of an LSM is shown in Figure \ref{fig:LSM}.
Input neurons are connected to a random subset of the liquid neurons.
The readout neurons may be connected to all the neurons in the liquid or a subset of them.
Connections between neurons in the liquid are based on probabilistic models of brain connectivity \cite{Maass2002_NeuralComput}:
\begin{equation}
P_{connection} (N_1, N_2) = q \cdot e^{-\frac{\mathbb{E}(N_1, N_2)}{r^2}}
\label{eq:probCon}
\end{equation}
where $N_1$ and $N_2$ represent two neurons and $\mathbb{E}(N_1, N_2)$ is the Euclidean distance between $N_1$ and $N_2$.
The variables $q$ and $r$ are two chosen constants.
In this paper, we use a 3-dimensional grid to define the positions of  neurons on the liquid.
The liquid functions as a temporal kernel, casting the input data into a higher dimension.
The LIF neurons allow for temporal state to be carried from one time step to another.
LSMs avoid the problem of training recurrent neural models by only training the synaptic weights from the liquid to the readout nodes, similar to extreme machine learning that use a random non-recurrent neural network for non-temporal data \cite{Huang2006_Neurocomputing}.
It is assumed that all temporal integration is encompassed in the liquid.
Thus, the liquid in an LSM acts similarly to the kernel in a support vector machine on streaming data by employing a temporal kernel.
In general, the weights and connections in the liquid do not change, although some studies have looked at plasticity in the liquid \cite{Norton2010_Neurocomputing}.

The readout neurons are the only neurons that have plastic synapses, allowing for synaptic weight updates via training.
Using each neurons firing state from the liquid, the temporal aspect of learning on temporal data is transformed to a static (non-temporal) learning problem.
As all temporal integration is done in the liquid, no additional mechanisms are needed to train the readout neurons.
Any classifier can be used, but often a linear classifier is sufficient.
Training of the readout neurons can be done in a batch or on-line manner \cite{Zhang2015_TNNLS}.

LSMs have been successfully applied to several applications including speech recognition \cite{Zhang2015_TNNLS}, vision \cite{Burgsteiner2007_AppInt}, and cognitive neuroscience \cite{Maass2002_NeuralComput, Buonomano2009_NatureReviews}.
Practical applications suffer from the fact that traditional LSMs take input in the form of spike trains.
Transforming numerical input data into spike data, such that the non-temporal data is represented temporally, is nontrivial.

\begin{table}[!t]
\renewcommand{\arraystretch}{1.3}
\caption{Parameters for the synapses (or connections between neurons) in the liquid.}
\label{table:liquidParams}
\centering
\begin{tabular}{|l|l c|}
\hline
Parameter & type & value\\
\hline
$r$ from Equation \ref{eq:probCon} & ALL & 2\\
\hline
\multirow{4}{*}{$q$ from Equation \ref{eq:probCon}} & $E \rightarrow  E$ & 0.45\\
 & $E \rightarrow I$ & 0.30\\
 & $I \rightarrow E$ & 0.60\\
 & $I \rightarrow  I$ & 0.15\\
\hline
\multirow{4}{*}{Synaptic weight} & $E \rightarrow  E$ & 3\\
 & $E \rightarrow I$ & 6\\
 & $I \rightarrow E$ & -2\\
 & $I \rightarrow  I$ & -2\\
\hline
\end{tabular}
\vskip -1mm
\end{table}

\section{Mapping the LSM onto the STPU}
\label{section:mapping}
In this section, we implement the LSM on the STPU.
There have been previous implementations of LSMs on hardware, however, in most cases an FPGA or VLSI chip has been designed specifically for a hardware implementation of an LSM.
Roy et al. \cite{Roy2014_IEEE_BCS} and also Zhang et al. \cite{Zhang2015_TNNLS} present a low-powered VLSI hardware implementation of an LSM.
Schrauwen et al. \cite{Schrauwen2008_NN} implement an LSM on an FPGA chip.
In contrast to other work, the STPU has been developed to be a general neuromorphic architecture. 
Other neuroscience work and algorithms are being developed against the STPU such as spike sorting and using spikes for median filtering \cite{Verzi2016_NN}.
Currently, we have an STPU simulator implemented in MATLAB as well as an implementation on an FGPA chip.
The MATLAB simulator has a one-to-one correspondence with the hardware implementation.

Given the constructs provided by the STPU, the LSM with a liquid composed of LIF neurons maps naturally onto the STPU.
We use the second-order synaptic response function of Equation \ref{eq:secondOrder} that is based on the work of Zhang et al. \cite{Zhang2015_TNNLS}.
They found that the second-order response function produced more dynamics in the liquid allowing the neural signals to persist longer after the input sequence had finished.
This lead to improved classification results.
Following Zhang et al., the synaptic properties of the liquid, including parameters for the connection probabilities between the liquid neurons defined in Equation \ref{eq:probCon} and the synaptic weights, are given in Table \ref{table:liquidParams}.
There are two types of neurons: excitatory ($E$) and inhibitory ($I$).
As has been observed in the brain \cite{Maass2002_NeuralComput}, the liquid is made up of an 80/20 network where 80\% of the nuerons are excitatory and 20\% of the neurons are inhibitory.
The probability of a synapse existing between two neurons and the weights between the neurons is dependent on the types of the considered neurons.
$E/I \rightarrow E/I$ denotes the presynaptic and postsynaptic neurons being connected by the synapse.
For example, $E \rightarrow I$ denotes the connection between an excitatory presynaptic neuron with an inhibitory postsynaptic neuron.
Excitatory neurons increase the action potential at a target neurons (positive synaptic weights) while the inhibitory neurons decrease the action potential (negative synaptic weights).
When the connections are generated between neurons in the liquid, the neurons are randomly connected according to Equation \ref{eq:probCon} with the parameters given in Table \ref{table:liquidParams}.
Each input neuron is randomly connected to a subset of 30\% of the neurons in the liquid with a weight of 8 or -8 chosen uniformly at random.

To implement the second-order synaptic response function, Equation \ref{eq:secondOrder} is sampled at discrete time steps and multiplied by the synaptic weight value between the neurons as specified in Table \ref{table:liquidParams}.
The discretely sampled weights are then encoded via multiple weights at corresponding cells in the temporal buffer for the postsynaptic neuron.
In this implementation, there is no synaptic delay ($\Delta_{kl}$ = 0) and $\tau^s_1$ is set to 4 and $\tau^s_2$ is set to 8 for excitatory neurons.
For inhibitory neurons, $\tau^s_1$ and $\tau^s_2$ are set to 4 and 2 respectively.
For all neurons, $\tau_j$ is set to 32.

The plastic readout neurons are connected to all of the neurons in the liquid.
Training is done off-line using a linear classifier on the average firing rate of the neurons in the liquid.
We examine the effect of the various linear classifiers below.

\subsection{Experiments}
To evaluate the effect of different parameters for the liquid state machine, we use a data set for spoken digit recognition of Arabic digits from 0 to 9 \cite{Hammami2010_ICCSIT}.
The dataset is composed of the time series Mel-frequency cepstral coefficients (MFCCs) of 8800 utterances of each digit from 88 speakers on with 10 repetitions per digit ($10 \times 10 \times 88$).
The MFCCs were taken from 44 male and 44 female native Arabic speakers between the ages of 18 and 40.
The dataset is partitioned into a training set from 66 speakers and a test set from the other 22 speakers.
We scale all variables between 0 and 1.

To evaluate the performance of the LSM, we examine the classification accuracy on the test set, and measure the separation in the liquid from the training set. 
If there is good separation within the liquid, then the state vectors from the trajectories for each class {\it should} be distinguishable from each other.
To measure the separability of a liquid $\Psi$ on a set of state vectors $O$ from the liquid perturbed by a given input sequence, we follow the definition from Norton and Ventura \cite{Norton2010_Neurocomputing}:
$
Sep(\Psi,O) = \frac{c_d}{c_v + 1}
$
where $c_d$ is the inter-class distance and $c_v$ is the intra-class variance.
Separation is the ratio of the distance between the classes divided by the class variance.

The inter-class difference is the mean difference of the center of mass for every pair of classes:
$
c_d = \sum_{l=1}^n \sum_{m=1}^n \frac{\| \mu(O_l) - \mu(O_m)\|_2}{n^2}
$
where $\|\cdot\|_2$ is the $L_2$ norm, $n$ is the number of classes, and $\mu(O_l)$ is the center of mass for each class. 
For a given class, the intra-class variance is the mean variance of the state vectors from the inputs from the center of mass for that class:
$
c_v = \frac{1}{n} \sum_{l=1}^n \frac{\sum_{o_k \in O_l}\|\mu(O_l)-o_k\|_2}{|O_l|}.
$

We investigate various properties of the liquid state machine, namely the synaptic response function, the input encoding scheme, the liquid topology, and the readout training algorithm. 
We also consider the impact of $\theta_j$ on the liquid.
A neuron $j$ will spike if $v_j$ exceeds $\theta_j$.
Thus, $\theta_j$ can have a significant impact on the dynamics of the liquid.
Beginning with a base value of 20 (as was used by Zhang et al.) we consider the effects of decreasing values of $\theta_j \in \{20, 17.5, 15, 12.5, 11, 10, 9, 7.5, 5, 3, 2.5, 2, 1\}$.
For default parameters, we use a reservoir of size $3\times 3\times 15$, we feed the the magnitude of the inputs into the input neurons (current injection) and a linear SVM to train the synapses for the readout neurons.

\label{section:experiments}
\subsubsection{Synaptic Response Functions}

\begin{table}[!t]
	\renewcommand{\arraystretch}{1.3}
	\caption{Separation values, average spiking rates, and classification accuracy from different synaptic response functions.
		}
	\label{table:responseFunction}
	\centering
	\begin{tabular}{|l|ccccc|}
		\hline
		Synaptic Res & TrainSep & TrainRate & TestSep & TestRate & SVM \\
		\hline
		Dirac Delta & 0.129 & 0.931 & 0.139 & 0.931 & 0.650 \\
		First-Order & 0.251 & 0.845 & 0.277 & 0.845 & 0.797 \\
		Second-Order & \textbf{0.263} & \textbf{0.261} & \textbf{0.290} & \textbf{0.255} & \textbf{0.868} \\
		\hline\hline
		First-Order 30 & 0.352 & 0.689 & 0.389 & 0.688 & 0.811 \\
		First-Order 40 & 0.293 & 0.314 & 0.337 & 0.314 & 0.817 \\
		First-Order 50 & 0.129 & 0.138 & 0.134 & 0.138 & 0.725 \\
		\hline
	\end{tabular}
	\vskip -1mm
\end{table}

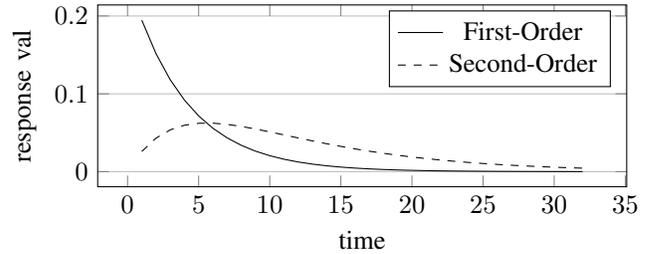
\begin{figure}
\begin{tikzpicture}
\begin{axis}[
width=0.475\textwidth,
height=4cm,
ymajorgrids,
x tick label style={/pgf/number format/1000 sep=},
ytick scale label code/.code={$\times$},
legend pos=north east,
xlabel=time,
ylabel=response val
]
\addplot[no marks] table[x=time,y=first] {data.csv};\addlegendentry{First-Order}
\addplot[no marks, dashed] table[x=time,y=second] {data.csv};\addlegendentry{Second-Order}
\end{axis}
\end{tikzpicture}
\caption{Visualization of the first- and second-order response functions.}
\label{fig:responseFunction}
\vskip -3mm
\end{figure}

We first investigate the effect of the synaptic response function using default parameters.
Using $\theta_j = 20$, the average separation values, average spiking rates and the classification accuracy of a linear SVM are given in Table \ref{table:responseFunction}.
As highlighted in bold, the second-order synaptic function (Equation \ref{eq:secondOrder}) achieves the largest separation values for training and testing, the lowest average spike rate, and the highest classification accuracy.
The average spike rate is significantly higher for the first-order response function than for the second-order response function, which is counter-intuitive since the second-order response function perpetuates the signal through the liquid longer.
However, examining the first-order and second-order response functions, as shown in Figure \ref{fig:responseFunction}, shows that the first-order response function has a larger initial magnitude and then quickly subsides.
The second-order response function has a lower initial magnitude, but is slower to decay giving a more consistent propagation of the spike through time.

Adjusting the value of $\theta_j$ to 30, 40, and 50 accommodate this behavior for the first-order response function (the bottom three rows int he Table \ref{table:responseFunction}) shows that an improvement can be made in the separation values, spiking rate, and classification accuracy.
Despite this improvement, the first-order response function does not achieve a better performance than the second-order response function for the classification accuracy.
The first-order response function does get a better separation score, but this does not translate into better accuracy.

\subsubsection{Input Encoding Schemes}

In traditional LSMs, the input is temporally encoded in the form of a spike train.
Unfortunately, most datasets are not temporally encoded, but rather are numerically encoded.
A spike train input aligns with neuroscience, but practically it is non-trivial to encode all information temporally as the brain does.
Therefore, we examine three possible encoding schemes:
1) rate encoding where the magnitude of the numeric value is converted into a rate and a spike train at that rate is fed into the liquid
2) bit encoding where the magnitude of the numeric value is converted into its bit representation at a given precision, and
3) current injection.
Rate encoding requires $n$ time steps to encode a single input by converting a magnitude to a rate. This is similar to binning and has some information loss.
Bit encoding, only requires one time step, however, it requires $m$ inputs per standard input to convert the magnitude into its $m$ bit-precise representation.
We set $m$ to 10.
Compared to current injection, the execution time increases linearly in the number of time steps for rate encoding.

\begin{table}[!t]
\renewcommand{\arraystretch}{1.3}
\caption{The separation values, average spiking rates of the liquid, and classification accuracy examining liquid topologies with different input encoding schemes and values for $\theta_j$.
The largest separation values and accuracies for each encoding scheme are in bold}
\label{table:encoding}
\centering
\begin{tabular}{|l|cccccc|}
\hline
 & \multicolumn{6}{|c|}{$\theta_j$}\\
Encoding Scheme & 20 & 15 & 10 & 5.5 & 3 & 2 \\
\hline
\hline
\multirow{3}{*}{Current Injection} & 0.263 & {\bf 0.409} & 0.378 & 0.334 & 0.324 & 0.324 \\ 
 & 0.261 & 0.580 & 0.750 & 0.843 & 0.873 & 0.878 \\
 & 0.868 & {\bf 0.905} & 0.894 & 0.873 & 0.868 & 0.866 \\
\hline
\multirow{3}{*}{Bit Encoding} & 0.271 & 0.310 & 0.338 & 0.350 & {\bf 0.353} & 0.357 \\
 & 0.434 & 0.497 & 0.544 & 0.592 & 0.634 & 0.620 \\
 & 0.741 & 0.741 & 0.735 & 0.755 & {\bf 0.764} & 0.761 \\
\hline
\multirow{3}{*}{Rate Encoding} & 0.164 & 0.364 & {\bf 0.622} & 0.197 & 0.047 & 0.048\\
& 0.146 & 0.199 & 0.594 & 0.952 & 0.985 & 0.985\\
& {\bf 0.747} & 0.733 & 0.643 & 0.601 & 0.548 & 0.558 \\
\hline
\end{tabular}
\vskip -1mm
\end{table}

Table \ref{table:encoding} shows the separation values (first row for each encoding scheme), average spiking rates (second row), and the accuracy from a linear SVM on the test set (third row) for the input encoding schemes with various values of $\theta_j$. 
The average spiking rate gives the percentage of neurons that were firing in the liquid through the time series and provides insight into how sparse the spikes are within the liquid.
Table \ref{table:encoding} shows a representative subset of the values that were used for $\theta_j$.
The bold values represent the highest separation value and classification accuracy for each encoding scheme. 

The results show that the value of $\theta_j$ has a significant effect on the separation of the liquid as well as the classification accuracy from the SVM.
This is expected as the dynamics of the liquid are dictated by when neurons fire.
A lower threshold allows for more spikes as is indicated in the increasing values of the average spiking rates as the values for $\theta_j$ decrease.
Overall, using rate encoding produces the greatest values in separation.
However, there is significant variability as the values for $\theta_j$ change.
For rate encoding, the greatest accuracy from the SVM is achieved with a low separation value.
In the other encoding schemes, separation and classification accuracy appear to be correlated.
The greatest classification accuracy is achieved from current injection.

\subsubsection{Liquid Topology}

The topology of the liquid in an LSM determines the size of the liquid and influences the connections within the liquid as the distance between the neurons impacts the connections made between neurons.
A more cubic liquid (e.g. $5 \times 5 \times 5$) should be more densely connected compared to a column of liquid (e.g. $2 \times 2 \times 20)$.
In this section, we examine using liquids with grids of  $3 \times 3 \times 15$,  $2 \times 2 \times 20$,  $4 \times 5 \times 20$, and  $5 \times 5 \times 5$. 
As before, we consider different values for $\theta_j$.
The separation values, average spike rates, and the accuracy from a linear SVM are given in Table \ref{table:topology} for the values of $\theta_j$ that provided the largest separation values for each topology configuration and encoding scheme combination.

\begin{table}[!t]
\renewcommand{\arraystretch}{1.3}
\caption{The separation values and average spike rate of the liquid using different liquid topologies.
	The largest separation values and accuracies for each topology are in bold}
\label{table:topology}
\centering
\begin{tabular}{|l|cc|cc|cc|}
\hline
Topology/ & \multicolumn{6}{c|}{Encoding Scheme}\\
Num Neurons&  \multicolumn{2}{c}{Current Injection} & \multicolumn{2}{c}{Bit Encoded} & \multicolumn{2}{c|}{Rate Encoded} \\
\hline
\hline
 \multicolumn{1}{|r|}{$\theta_j$:} & 15 & 12.5 & 3 & 2 & 11 & 10 \\
\hline
\multirow{3}{*}{3x3x15/135} & 0.409 & 0.405 & 0.353 & 0.357 & {\bf 0.622} & {\bf 0.622}\\
 & 0.580 & 0.693 & 0.634 & 0.620 & 0.563 & 0.594\\
 & {\bf 0.905} & 0.884 & 0.764 & 0.761 & 0.658 & 0.643 \\
\hline
\multicolumn{1}{|r|}{$\theta_j$:} & 12.5 & 11 & 3 & 2 & 10 & 9 \\
\hline
\multirow{3}{*}{5x5x5/125} & {\bf 0.467} &  0.432 & 0.384 & 0.380 & 0.384 & 0.451 \\
 & 0.534 & 0.682 & 0.596 & 0.586& 0.473 & 0.600 \\
 & 0.889 & {\bf 0.900} & 0.765 & 0.755 & 0.637 & 0.610 \\
\hline
 \multicolumn{1}{|r|}{$\theta_j$:} & 15 & 12.5 & 15 & 12.5 & 15 & 12.5\\
\hline
\multirow{3}{*}{4x5x10/200} & 0.418 & 0.384  & {\bf 0.506} & 0.501 & 0.479 & 0.352\\
& 0.698 & 0.760 & 0.576 & 0.603 & 0.289 & 0.577\\
 & {\bf 0.879} & 0.876 & 0.830 & 0.835 & 0.644 & 0.612 \\
\hline
 \multicolumn{1}{|r|}{$\theta_j$:} & 11 & 10 & 3 & 2 & 10 & 9 \\
\hline
\multirow{3}{*}{2x2x20/80} & {\bf 0.397} & 0.389 & 0.264 & 0.247 & 0.305 & 0.255 \\
 & 0.608 & 0.658 & 0.537 & 0.550 & 0.609 & 0.846 \\
 & 0.903 & {\bf 0.912} & 0.672 & 0.666 & 0.695 & 0.678 \\
\hline
\end{tabular}
\vskip -1mm
\end{table}

Again, the value for $\theta_j$ has a significant impact on the separation of the liquid and the classification accuracy.
The greatest separation values and classification accuracies for each topology is highlighted in bold.
For all of the topologies, current injection achieves the highest classification accuracy.
Interestingly, the separation values across encoding schemes and topologies do not correlate with accuracies. 
Within the same encoding scheme and topology, however, the accuracy generally improves as the separation increases.

For current injection, the different topologies do not appear to have a significant impact on the classification accuracy except for the $4 \times 5 \times 10$ topology which has a decrease in accuracy.
This may be due to increased number of liquid nodes that are used as input to the SVM.
The converse is true for bit encoding as the $4 \times 5 \times 10$ topology achieves the highest accuracy possibly due to the increased number of inputs due to the bit representation of the input.

\subsubsection{Readout Training Algorithms}

How the plastic synapses are trained will have a significant effect on the performance of the LSM. 
Traditionally, LSMs use a linear classifier based on the assumption that the liquid has transformed the state space such that the problem is linearly separable.
Linear models are represented as a set of weights and a threshold---which can be implemented in neuromorphic hardware.
By using a linear model, the liquid and the classification can all be done on the STPU avoiding the overhead of going off chip to make a prediction.
We consider four linear classifiers: 1) linear SVM, 2) linear discriminant analysis (LDA), 3) ridge regression, and 4) logistic regression.
With each of these algorithms, we use the default parameters as they are set in the Statistics and Machine Learning Toolbox in MATLAB. 

We examine the classification of each of the linear classifiers on the topologies and $\theta_j$ values that achieved the highest classification accuracy on the linear SVM in previous experiments.
We also limit ourselves to examining current injection for the input scheme as current injection consistently achieved the highest classification accuracy.
The results are shown in Table \ref{table:classifiers}.
LDA consistently achieves the highest classification accuracy of the considered classifiers.
The highest classification accuracy achieved is 0.946.

\begin{table}[!t]
	\renewcommand{\arraystretch}{1.3}
	\caption{Classification accuracy on the test set from different linear classifiers.
		The greatest accuracy for each topology is in bold}
	\label{table:classifiers}
	\centering
	\begin{tabular}{|l|c|c|c|c|}
		\hline
		Linear Model: & $3 \times 3 \times 15$ & $5 \times 5 \times 5$ & $4 \times 5 \times 10$ & $2 \times 2 \times 20$\\
		& $\theta_j$ = 15  & $\theta_j$ = 11 & $\theta_j$ = 15 & $\theta_j$ = 10 \\
		\hline
		Linear SVM & 0.906 & 0.900 & 0.900 & 0.914 \\
		LDA & {\bf 0.921} & {\bf 0.922} & {\bf 0.922} & {\bf 0.946}\\
		Ridge Regress & 0.745 & 0.717 & 0.717 & 0.897\\
		Logistic Regress & 0.431 & 0.254 & 0.254 & 0.815\\
		\hline
	\end{tabular}
\vskip -1mm
\end{table}



\section{Conclusion and Future Work}
\label{section:conc}

In this paper, we presented the {\it Spiking Temporal Processing Unit} or STPU---a novel neuromorphic processing architecture.
It is well suited for efficiently implementing neural networks and synaptic response functions of arbitrary complexity.
This is facilitated by using the temporal buffers associated with each neuron in the architecture. 
The capabilities of the STPU, including complex synaptic response functions, were demonstrated through implementing the functional mapping and implementation of an LSM onto the STPU architecture.

As neural algorithms grow in scale and conventional processing units reach the limits of Moore's law, neuromorphic computing architectures, such as the STPU, allow efficient implementations of neural algorithms.
However, neuromorphic hardware is based on spiking neural networks to achieve low energy.
Thus, more research is needed to understand and develop spiking-based algorithms.
\bibliographystyle{IEEEtran}
\bibliography{IEEEabrv,bibtex}

 \newcommand{\noop}[1]{}
\begin{thebibliography}{10}
\providecommand{\url}[1]{#1}
\csname url@samestyle\endcsname
\providecommand{\newblock}{\relax}
\providecommand{\bibinfo}[2]{#2}
\providecommand{\BIBentrySTDinterwordspacing}{\spaceskip=0pt\relax}
\providecommand{\BIBentryALTinterwordstretchfactor}{4}
\providecommand{\BIBentryALTinterwordspacing}{\spaceskip=\fontdimen2\font plus
\BIBentryALTinterwordstretchfactor\fontdimen3\font minus
  \fontdimen4\font\relax}
\providecommand{\BIBforeignlanguage}[2]{{%
\expandafter\ifx\csname l@#1\endcsname\relax
\typeout{** WARNING: IEEEtran.bst: No hyphenation pattern has been}%
\typeout{** loaded for the language `#1'. Using the pattern for}%
\typeout{** the default language instead.}%
\else
\language=\csname l@#1\endcsname
\fi
#2}}
\providecommand{\BIBdecl}{\relax}
\BIBdecl

\bibitem{Deng2013_ICASSP}
L.~Deng, G.~E. Hinton, and B.~Kingsbury, ``New types of deep neural network
  learning for speech recognition and related applications: an overview,'' in
  \emph{Proceedings of International Conference on Acoustics, Speech, and
  Signal Processing (ICASSP)}, 2013, pp. 8599--8603.

\bibitem{Ciresan2012_CVPR}
D.~Ciresan, U.~Meier, and J.~Schmidhuber, ``Multi-column deep neural networks
  for image classification,'' in \emph{Proceedings of the 2012 IEEE Conference
  on Computer Vision and Pattern Recognition}.\hskip 1em plus 0.5em minus
  0.4em\relax IEEE Computer Society, 2012, pp. 3642--3649.

\bibitem{Socher2013_EMNLP}
R.~Socher, A.~Perelygin, J.~Wu, J.~Chuang, C.~D. Manning, A.~Ng, and C.~Potts,
  ``Recursive deep models for semantic compositionality over a sentiment
  treebank,'' in \emph{Proceedings of the 2013 Conference on Empirical Methods
  in Natural Language Processing}.\hskip 1em plus 0.5em minus 0.4em\relax
  Association for Computational Linguistics, October 2013, pp. 1631--1642.

\bibitem{Backus1978_CommACM}
J.~Backus, ``Can programming be liberated from the von neumann style?: A
  functional style and its algebra of programs,'' \emph{Communications of the
  ACM}, vol.~21, no.~8, pp. 613--641, Aug. 1978.

\bibitem{Follett2009_JofChildNeur}
P.~L. Follett, C.~Roth, D.~Follett, and O.~Dammann, ``White matter damage
  impairs adaptive recovery more than cortical damage in an in silico model of
  activity-dependent plasticity,'' \emph{Journal of Child Neurology}, vol.~24,
  no.~9, pp. 1205--1211, 2009.

\bibitem{Sejnowski1995_Nature}
T.~J. Sejnowski, ``{Time for a new neural code?}'' \emph{Nature}, vol. 376, no.
  6535, pp. 21--22, Jul. 1995.

\bibitem{Merolla2014_Science}
P.~Merolla, J.~Arthur, R.~Alvarez-Icaza, A.~Cassidy, J.~Sawada, F.~Akopyan,
  B.~Jackson, N.~Imam, C.~Guo, Y.~Nakamura, B.~Brezzo, I.~Vo, S.~Esser,
  R.~Appuswamy, B.~Taba, A.~Amir, M.~Flickner, W.~Risk, R.~Manohar, and
  D.~Modha, ``A million spiking-neuron integrated circuit with a scalable
  communication network and interface,'' \emph{Science}, pp. 668--673, August
  2014.

\bibitem{Furber2013_TransOnComp}
S.~B. Furber, D.~R. Lester, L.~A. Plana, J.~D. Garside, E.~Painkras, S.~Temple,
  and A.~D. Brown, ``Overview of the spinnaker system architecture,''
  \emph{IEEE Transactions on Computers}, vol.~62, no.~12, pp. 2454--2467, 2013.

\bibitem{Schemmel2008_IJCNN}
J.~Schemmel, J.~Fieres, and K.~Meier, ``Wafer-scale integration of analog
  neural networks,'' in \emph{IEEE International Joint Conference on Neural
  Networks}, june 2008, pp. 431 --438.

\bibitem{Zhang2015_TNNLS}
Y.~Zhang, P.~Li, Y.~Jin, and Y.~Choe, ``A digital liquid state machine with
  biologically inspired learning and its application to speech recognition,''
  \emph{{IEEE} Transactions on Neural Networks and Learning Systems}, vol.~26,
  no.~11, pp. 2635--2649, 2015.

\bibitem{Maass2002_NeuralComput}
W.~Maass, T.~Natschl\"{a}ger, and H.~Markram, ``Real-time computing without
  stable states: A new framework for neural computation based on
  perturbations,'' \emph{Neural Computation}, vol.~14, no.~11, pp. 2531--2560,
  Nov. 2002.

\bibitem{Verzi2017_IJCNN}
S.~J. Verzi, C.~M. Vineyard, E.~D. Vugrin, M.~Galiardi, C.~D. James, and J.~B.
  Aimone, ``Optimization-based computation with spiking neurons,'' in
  \emph{Proceedings of the IEEE International Joint Conference on Neural
  Network}, 2017, p. Accepted.

\bibitem{Verzi2016_NN}
S.~J. Verzi, F.~Rothganger, O.~D. Parekh, T.-T. Quach, N.~E. Miner, C.~D.
  James, and J.~B. Aimone, ``Computing with spikes: The advantage of
  fine-grained timing,'' \noop{2016}submitted.

\bibitem{Dayan2001_TheoreticalNeuroSci}
P.~Dayan and L.~F. Abbott, \emph{Theoretical neuroscience : computational and
  mathematical modeling of neural systems}, ser. Computational
  neuroscience.\hskip 1em plus 0.5em minus 0.4em\relax Cambridge (Mass.),
  London: MIT Press, 2001.

\bibitem{Benjamin2014_IEEE}
B.~V. Benjamin, P.~Gao, E.~McQuinn, S.~Choudhary, A.~Chandrasekaran, J.-M.
  Bussat, R.~Alvarez-Icaza, J.~Arthur, P.~Merolla, and K.~Boahen, ``Neurogrid:
  A mixed-analog-digital multichip system for large-scale neural simulations,''
  \emph{Proceedings of the IEEE}, vol. 102, no.~5, pp. 699--716, 2014.

\bibitem{Schemmel2010_SymposiumCircuitSystems}
J.~Schemmel, D.~Briiderle, A.~Griibl, M.~Hock, K.~Meier, and S.~Millner, ``A
  wafer-scale neuromorphic hardware system for large-scale neural modeling,''
  in \emph{Proceedings of 2010 IEEE International Symposium on Circuits and
  Systems}, May 2010, pp. 1947--1950.

\bibitem{Furber2014_IEEE}
S.~Furber, F.~Galluppi, S.~Temple, and L.~A. Plana, ``The spinnaker project,''
  \emph{Proceedings of the {IEEE}}, vol. 102, no.~5, pp. 652--665, 2014.

\bibitem{Severa2016_NIPS_WS}
W.~Severa, K.~D. Carlson, O.~Parekh, C.~M. Vineyard, and J.~B. Aimone, ``{Can
  we be formal in assessing the strengths and weaknesses of neural
  architectures? A case study using a spiking cross-correlation algorithm},''
  in \emph{NIPS Workshop ‘Computing with Spikes’}, 2016.

\bibitem{Lukosevicius2009_CSReview}
M.~Luko{\v s}evi{\v c}ius and H.~Jaeger, ``Reservoir computing approaches to
  recurrent neural network training,'' \emph{Computer Science Review}, vol.~3,
  no.~3, pp. 127--149, August 2009.

\bibitem{Jaeger2003_NIPS}
H.~Jaeger, ``Adaptive nonlinear system identification with echo state
  networks,'' in \emph{Advances in Neural Information Processing Systems
  15}.\hskip 1em plus 0.5em minus 0.4em\relax MIT Press, 2003, pp. 609--616.

\bibitem{Huang2006_Neurocomputing}
G.~B. Huang, Q.~Y. Zhu, and C.~K. Siew, ``{Extreme learning machine: theory and
  applications},'' \emph{Neurocomputing}, vol.~70, no. 1-3, pp. 489--501, 2006.

\bibitem{Norton2010_Neurocomputing}
D.~Norton and D.~Ventura, ``Improving liquid state machines through iterative
  refinement of the reservoir,'' \emph{Neurocomputing}, vol.~73, no. 16-18, pp.
  2893--2904, 2010.

\bibitem{Burgsteiner2007_AppInt}
H.~Burgsteiner, M.~Kr{\"o}ll, A.~Leopold, and G.~Steinbauer, ``Movement
  prediction from real-world images using a liquid state machine,''
  \emph{Applied Intelligence}, vol.~26, no.~2, pp. 99--109, 2007.

\bibitem{Buonomano2009_NatureReviews}
D.~Buonomano and W.~Maass, ``State-dependent computations: Spatiotemporal
  processing in cortical networks,'' \emph{Nature Reviews Neuroscience},
  vol.~10, no.~2, pp. 113--125, 2009.

\bibitem{Roy2014_IEEE_BCS}
S.~Roy, A.~Banerjee, and A.~Basu, ``Liquid state machine with dendritically
  enhanced readout for low-power, neuromorphic {VLSI} implementations,''
  \emph{IEEE Transactions on Biomedical Circuits and Systems}, vol.~8, no.~5,
  2014.

\bibitem{Schrauwen2008_NN}
B.~Schrauwen, M.~D`Haene, D.~Verstraeten, and D.~Stroobandt, ``Compact hardware
  liquid state machines on fpga for real-time speech recognition,''
  \emph{Neural Networks}, no.~21, pp. 511--523, 1 2008.

\bibitem{Hammami2010_ICCSIT}
N.~Hammami and M.~Bedda, ``Improved tree model for arabic speech recognition,''
  in \emph{Proceddings of the IEEE International Conference on Computer Science
  and Information Technology}, 2010.

\end{thebibliography}
%
%
%

\end{document}